# Automated turnkey microcomb for low-noise microwave synthesis

Kunpeng Jia[1]†*, Xinwei Yi[1]†, Xiaohan Wang[1], Yunfeng Liu[2], Shu-Wei Huang[3], Xiaoshun Jiang[1], Wei Liang[2]*, Zhenda Xie[1]*, Shi-ning Zhu[1]

[1] *National Laboratory of Solid State Microstructures, School of Electronic Science and Engineering, College of Engineering and Applied Sciences, School of Physics, and Collaborative Innovation Center of Advanced Microstructures, Nanjing University, Nanjing 210093, China.*

[2] *Suzhou Institute of Nano-tech and Nano-bionics, Chinese Academy of Sciences, Suzhou 215123, China.*

[3] *Department of Electrical, Computer and Energy Engineering, University of Colorado Boulder, Boulder, Colorado 80309, USA.*

* E-mail: jiakunpeng@nju.edu.cn; wliang2019@sinano.ac.cn; xiezhenda@nju.edu.cn

† These authors contributed equally to this work.

**Microresonator-based optical frequency comb (microcomb) has the potential to revolutionize the accuracy of frequency synthesizer in radar and communication applications. However, fundamental limit exists for low noise microcomb generation, especially in low size, weight, power and cost (SWaP-C) package. Here we resolve this limit, by the demonstration of an automated turnkey microcomb, operating close to its low quantum-limited phase noise, within a compact setup size of 85 mm × 90 mm × 25 mm. High quality factor fiber Fabry-Perot resonator (FFPR), with Q up to $4.0\times10^9$, is the key for both low quantum noise and pump noise limit, in the diode-pump case in a self-injection locking scheme. Low phase noise of −80 and −105 dBc/Hz at 100 Hz, −106 and −125 dBc/Hz at 1 kHz, −133 and −148 dBc/Hz at 10 kHz is achieved at 10.1 GHz and 1.7 GHz repetition frequencies, respectively. With the simultaneous automated turnkey, low-noise and direct-diode-pump capability, our microcomb is ready to be used as a low-noise frequency synthesizer with low SWaP-C and thus field deployability.**

The optical frequency comb (OFC) keeps the highest precision in both time and frequency metrology[1-3]. Such high precision from the OFC may also revolutionize the accuracy in field-deployable devices, for massive applications including microwave frequency synthesis[4-7], optical clock[8,9], and astronomical observations[10-12] etc., and microresonator-based optical frequency comb (microcomb)[13-15] is the most promising candidate, to be portable with low size, weight, power and cost (SWaP-C). However, fundamental limit to the low noise exists for the conventional microcombs, and thus limits its application that requires superior precision, like the microwave frequency synthesis. On the one hand, the strong nonlinearity in small mode volume sets relatively high quantum noise limit[16, 17], and on the other hand, such quantum limit is not easily accessible in a normal microcomb, because of the technical noises like pump laser noise etc. Recent experiments have shown that the low quantum-limited phase noise can be achieved in resonators with relatively large mode volume, including fiber Fabry-Perot resonators (FFPRs)[18,19] and silica microdisks[20-22]. The drawback is the relatively high pump power that requires amplified lasers as pump, which results in a non-integrated setup. The other challenge is to operate the microcomb in a plug-and-play manner, which is known as the "turnkey operation" capability as required in practical application. In fact, the startup of the microcomb relies on complex nonlinear optical interaction[23-26], and the low noise soliton state can only be accessed with optimum initial state preparation and certain dynamic processes. Massive efforts have been devoted to the turnkey operation for the microcomb, and its self-starting operation has been demonstrated only recently in silicon nitride resonator in a self-injection scheme[27,28]. For the practical application of microcomb as a frequency synthesizer, the question is still open whether a microcomb can be low in quantum limited phase noise with turn-key capability, and in a compacted direct diode pump setup.

Here we report the first automated turnkey microcombs with low phase noise that is close to quantum limit, in an integrated self-injection-locking setup with direct diode pump. These microcombs are generated using FFPRs that we fabricate, with high quality factor (Q) measured up to $4.0\times10^9$, which is the key to enable direct diode pump and pump noise reduction. The system map preparation is carried out automatically by field programmable gate array (FPGA) for pump power and phase control for accessing single soliton state regime and its turnkey generation, so that a plug-and-play microcomb operation is achieved. We build our microcomb with different repetition frequencies at 10.1 GHz and 1.7 GHz, for radio frequency (RF) generation in X band and L band, respectively. Low phase noise of the generated 10.1 and 1.7 GHz RF signals is measured to be −80 and −105 dBc/Hz at 100 Hz, −106 and −125 dBc/Hz at 1 kHz, −133 and −148 dBc/Hz at 10 kHz, and −159 and −150 dBc/Hz at 10 MHz, respectively, which are close to the low quantum noise limit of FFPRs. Low Allan deviations are measured to be below $2 \times 10^{-11}$ and $3 \times 10^{-11}$ at 1 s, $7 \times 10^{-10}$ and $8 \times 10^{-10}$ at 100 s averaging time, respectively. FFPRs, pump laser diode and all the micro optical components are packaged in a hand-held size of 85 mm × 90 mm × 25 mm for easy usage and portability. Above results mark a major step of microcombs towards field-deployable low-noise microwave synthesizer, for radar[29,30] and communication[31,32] applications.

The noise performance of a microcomb is fundamentally limited by the quantum noise in a microresonator geometry, which is further affected by the technical noise in practical setup on top of the quantum noise. Our study shows that the quality factor of a microresonator plays an important role for the reduction of quantum noise limit, while the pump power is limited, as in the direct diode pump case. The noise from the pump can dominate in the technical noise, which limits the noise performance beyond the quantum-limit noise, especially in the low offset frequency regime. High Q may help to lower the pump noise limit in a self-injection locking setup.

For a given Kerr resonator, we assume a mode-locked soliton state is generated at certain ratio $\kappa$ for pump power $P_{\mathrm{in}}$ relative to parametric threshold power $P_{\mathrm{th}}$, so that the quantum-limited phase noise of a microcomb can be written as[17,33-35] (see details in Supplementary Information)

$$
\begin{aligned}
L_{\Phi} &= \frac{\sqrt{2}\pi^3\hbar c}{\lambda}\sqrt{\frac{1}{\kappa(-D)}}\bullet\frac{1}{P_{th}}, \quad f \gg \gamma \\
L_{\Phi} &= \frac{\sqrt{2}\pi^5\hbar c^3}{\lambda^3}\sqrt{\frac{-D}{\kappa}}\left[\frac{1}{96\kappa}+\frac{\kappa}{24\pi^2}\right]\frac{1}{f^2}\bullet\frac{1}{P_{th}Q^2}, \quad f \ll \gamma
\end{aligned}
\tag{1}
$$

where $D$, $\gamma$, $c$, $\lambda$, and $f$ are the dimensionless dispersion parameter, half width half maximum of the cavity resonance, speed of light in the vacuum, pump wavelength, and the offset radio frequency, respectively. While the quantum noise at high offset frequency is only related to the pump power, the noise at low offset frequency is inversely proportional to $Q^2$ at certain pump power. Indeed, phase noise at such low offset frequency plays an important role in the radio frequency synthesis, as $\gamma$ still covers the frequency of kilohertz with high Q of $10^9$.

On the other hand, pump noise can also be transferred to the microcomb noise, resulting in higher noise than the quantum limit. With the pump as a comb line, the lower limit of the microcomb repetition frequency noise is the frequency division from the pump noise[4,36], by the number of comb lines. In a self-injection locking setup, significant noise reduction can be achieved when the microresonator has much higher Q than that of the laser diode $Q_{\mathrm{LD}}$[37-39], and the reduction factor can be calculated as

$$
\eta = 1+\frac{Q}{Q_{LD}}\sqrt{\beta(1+\alpha_g^2)}, \tag{2}
$$

where $\beta$ and $\alpha_{\mathrm{g}}$ are the feedback strength and linewidth-broadening factor, respectively. So that it is also the key to improve the Q of the microresonator to achieve lower pump phase noise and thus the pump induced comb noise.

In order to achieve high Q for an FFPR, we develop fine polishing and high-reflectivity

(> 99.9%) coating technique. We prepare two FFPRs with different lengths of 10 mm and 59 mm (Figs. 1a and 1c), corresponding to repetition frequencies of 10.1 GHz and 1.7 GHz, respectively. Such frequencies are important for radar and communication applications. The Qs of the two FFPRs are characterized by cavity ring-down measurement[40]. By fast sweeping the frequency of a tunable external cavity diode laser (CTL 1550, Toptica) over the FFPRs resonances, the decay times are fitted to be 492 ns and 3308 ns, corresponding to Qs of $6.0\times10^8$ and $4.0\times10^9$ (Figs. 1b and 1d). Such high Q is about two orders of magnitude higher than that in our previous work[18], so that we can choose fiber with even larger mode volume and lower nonlinearity, for the FFPR fabrication and still get sub-100 mW threshold power for single soliton generation. In experiment, the fiber we use is few-mode fiber (YOFC), with fundamental mode field area of 112 $\mu m^2$ and small nonlinear index ($n_2 = 3\times10^{-20}$ $m^2/W$).

It is interesting to compare the quantum limit and pump noise limit in our FFPR, with results shown in Fig. 1e. The pump noise limit is calculated by dividing the phase noise of the injection-locked DFB laser to RF domain with a ratio of frequency division. The DFB laser phase noise data we use is from our experimental measurement. Taking the 1.7 GHz repetition FFPR for example, that only if the Q reaches the threshold value of $6.4 \times 10^9$, can it be possible to achieve quantum noise without being limited by the pump noise above the cutoff frequency of 10 Hz. Although the Q ($=4.0\times10^9$) of our FFPR is still not high enough to make the pump noise limit lower than quantum limit over the whole offset frequencies, it has been able to push crossing frequency down to 22 Hz, which is easy to be further suppressed by active feedback control. Figure 1f plots the difference between the pump noise limit and quantum noise at 10 Hz offset frequency as a function of FFPR Q and repetition frequency. The black solid line is formed by connecting all the 0 dB points and gives the threshold Q for different repetition frequencies.

Our microcomb setup is schematically shown in Fig. 1g. The pump light is from a distributed feedback (DFB) laser diode, and is directed through the FFPR, with part of the transmission feedback into the DFB laser diode, for self-injection locking[41]. Our simulation shows that self-starting single soliton state can be achieved when certain relative detuning between the injection-locked laser and the cavity resonance is achieved at certain pump power, as the balance for the cavity dispersion and nonlinear phase modulation. Like the cases in other works on microcombs, the single soliton zone is only a small portion in the system map of the pump power and detuning, and it is accessible by tuning the pump power and the feedback phase in the self-injection locking loop[27,42], as shown in Fig. 1h. In experiment, the whole setup is assembled in a temperature-controlled baseplate of 85 mm × 90 mm × 25 mm size, as shown in the inset of Fig. 1g. The pump power can be changed by tuning the pump current and a 0.8 mm-thick silicon sheet is inserted in the feedback path for the phase tuning by the temperature control using a resistive heater. The initial state preparation is automated using FPGA for the turnkey soliton generation.

We first measure the pump laser performance in the setup. Figures. 2a and 2b shows the single mode pump spectrum captured by using an optical spectral analyzer (OSA), and the pump output power as a function of injection current, respectively. We perform the frequency noise measurements of the pump in comparison to its non-locked free-running state, using the correlated delayed self-heterodyne method[43], as shown in Fig. 2c. Ultra-narrow fundamental linewidths of 72.2 mHz and 38.6 mHz are measured for DFB lasers injection-locked to 10.1 GHz and 1.7 GHz FFPRs, respectively. Compared to the 66.4 kHz fundamental linewidth of free-running DFB laser, a linewidth reduction factor over $10^6$ is achieved. The purified pump spectrum can be delivered to every comb line in following microcomb mode-locking, and found a good basis for the low noise comb generation.

Then we study the soliton comb formation dynamics in our setup. The cold laser frequency is tuned in relative to the cold FFPRs resonance, by the modulation of the pump current with a triangle wave signal. The output light of FFPRs is detected by a photodetector, with the intensity trace captured by an oscilloscope. The waveforms are shown in Figs. 2d and 2e. When the laser frequency is tuned into deep resonance, the power coupled into FFPRs exceeds the parametric oscillation threshold and boosts the four-wave mixing to chaotic comb state. With proper phase delay, the signature of soliton generation, soliton step, will subsequently appears when the pump is red detuned, and the comb evaluates into a passively stable, mode-locked and low noise soliton state. Due to the injection locking induced mode pulling, the soliton ranges are significantly extended to be 1.33 GHz and 0.22 GHz for 10.1 GHz and 1.7 GHz FFPRs, respectively, which are beneficial for the stable operation in the soliton states. Actually, the single soliton states can be easily accessed by slow and manual pump current tuning, which is consistent with our simulation for the self-starting soliton generation.

The self-starting soliton generation found the basis of a turnkey microcomb generation, but initial state preparation needs to be performed for the microcomb into a “turnkey” zone[27,42]. We use FPGA for the initial state preparation in our setup because of its powerful data processing, high data throughput and low latency capabilities, for the servo via the laser current and the phase tuning using the silicon sheet. The automated microcomb generation strategy is illustrated in Fig. 2f. It is a two-dimensional scan process for the turnkey soliton state, over the feedback phase and the pump current. This scan is under the control of FPGA, which communicates with the monitor photodiode and driving electronics through analog-to-digital converter (ADC) and digital-to-analog converter (DAC). With the feedback phase fixed at each value, the laser current is continuous modulated in a triangular wave, until the single soliton step is automatically identified from FFPR transmission signal by FPGA. Because of the low thermal mass of the silicon phase sheet, the whole two-dimensional scan process can be finished within tens of seconds. Then the system is in a “single soliton ready” state for stable soliton microcomb generation.

The spectra of generated single soliton states are recorded by an OSA, as shown in Figs. 3a and 3b. The dashed red curves show clean single soliton envelopes of $\mathrm{sech}^2$ fitting for soliton pulses widths of 583 fs and 954 fs for combs at 10.1 GHz and 1.7 GHz, respectively.

We perform the temporal autocorrelation measurement for 10.1 GHz comb and the trace is plotted in the left inset of Fig. 3a, showing 1 ps pulse duration, corresponding to an actual pulse width of about 650 fs, which is close to the result indicated from the optical spectrum. By sending the amplified single soliton combs with pump filtered into a fast photodetector, beatnote signals are captured on an electrical spectrum analyzer (ESA), revealing highly coherent mode-locked states generation as shown in Figs. 3c and 3d.

The single sideband (SSB) phase noise of the generated RF beatnote is characterized by a phase noise analyzer (APPH40G, Anapico), as the result shown in Figs. 4a and 4b. Of note, the generated 1.7 GHz signal has reached the quantum-limited phase noise in offset frequency range beyond 100 Hz. The cutoff frequency is pushed two orders of magnitude towards lower offset frequency compared to our previous work based on FFPR, which reached quantum-limited phase noise over 10 kHz. Such improvement can be contributed to the significant pump frequency noise reduction that lowers the achievable phase noise down to the quantum noise. The phase noise of 10.1 GHz signal is also close to quantum-limited noise especially for high offset frequency. The obtained phase noise for the 10.1 GHz repetition frequency in X band is −80 dBc/Hz at 100 Hz, −106 dBc/Hz at 1 kHz, and −133 dBc/Hz at 10 kHz. It is the first time that turnkey soliton microcomb can reach such low noise that is close to its quantum limit. Such quantum limit is already low using our high-Q FFPR, with relatively large mode volume and small nonlinear coefficient for a Kerr microresonator. In fact, being free-running currently, the performance of our microcomb is comparable to the best record of low-noise microwave using a soliton microcomb-based transfer oscillator with active feedback control, which features −135 dBc/Hz at 10 kHz offset frequency[44]. However, our scheme bypasses the requirement for sub-Hz-linewidth ultra-stable laser, sophisticated soliton initialization technique and complicated self-referencing operation. Also, to the best of our knowledge, this is the first compact microcomb-based low noise microwave synthesizer that is capable of reaching a frequency as low as 1.7 GHz.

Finally, we characterize the repetition frequency stability of the two comb sources by the Allan deviation measurement using a frequency counter with a 10 MHz Rubidium clock signal as time base. Of note, to improve the long-term stability of the generated single soliton state, the laser current is feedback controlled through the optional Proportional Integral (PI) loop of FPGA in Fig. 2f by using the variation of soliton power as an error signal. The Allan deviation measurement results are shown in Figs. 4c and 4d. The measured frequency instability of 10.1 GHz and 1.7 GHz signals is below $2 \times 10^{-11}$ and $3 \times 10^{-11}$ at 1 s, $7 \times 10^{-10}$ and $8 \times 10^{-10}$ at 100 s averaging time, respectively. For further fully stabilizing the microcomb, we test the tuning range of the repetition frequency. The 10.1 GHz signal can be changed in a frequency range of near 1 kHz (corresponding to $10^{-7}$ fraction of the carrier frequency) by simply tuning laser current. The frequency modulation efficiency is 1.3 kHz/mA. Such tunability is far enough for long-term repetition frequency calibration through feedback control of the laser current as the frequency instability is far below $10^{-7}$ in a time scale of 1000 s.

In conclusion, we have demonstrated the first turnkey microcombs with low phase noise that is close to quantum limit. This quantum noise limit is also low with our high-Q FFPR design and is accessed with significant pump noise reduction, at offset frequency down to 100 Hz for the first time. Full turnkey capability is presented, including both automated initial state preparation and self-starting single soliton generation, for the plug-and-play microcomb setup, within a handheld setup of 85 mm × 90 mm × 25 mm. As a result, low phase noises of −80 and −105 dBc/Hz at 100 Hz, −106 and −125 dBc/Hz at 1 kHz, −133 and −148 dBc/Hz at 10 kHz, and −159 and −150 dBc/Hz at 10 MHz are achieved for 10.1 and 1.7 GHz RF signals, respectively. We also test the tuning capability of the repetition rate using the direct modulation of drive current for the pump diode, and $10^{-7}$ frequency modulation can be achieved. Such modulation range is larger than the long-term stability of our microcomb, which can be used for active feedback locking to fully stabilize this frequency synthesizer. It is worth noting that the Q of our FFPR is still limited by the coating reflectivity, while the propagation loss limited Q is on the order of $10^{11}$, considering the fiber loss on the order of 0.2 dB/km. By improving the polishing and coating technique, higher Q can be expected, enabling even larger mode diameter and lower nonlinearity at the same pump requirement, and lower quantum noise limit is possible. With the current results and these potential improvements, our FFPR-based microcomb meets the fundamental requirements of a low-noise and field-deployable frequency synthesizer. With further stabilization, a portable frequency standard can be expected, for more demanding applications such like optical clock etc.

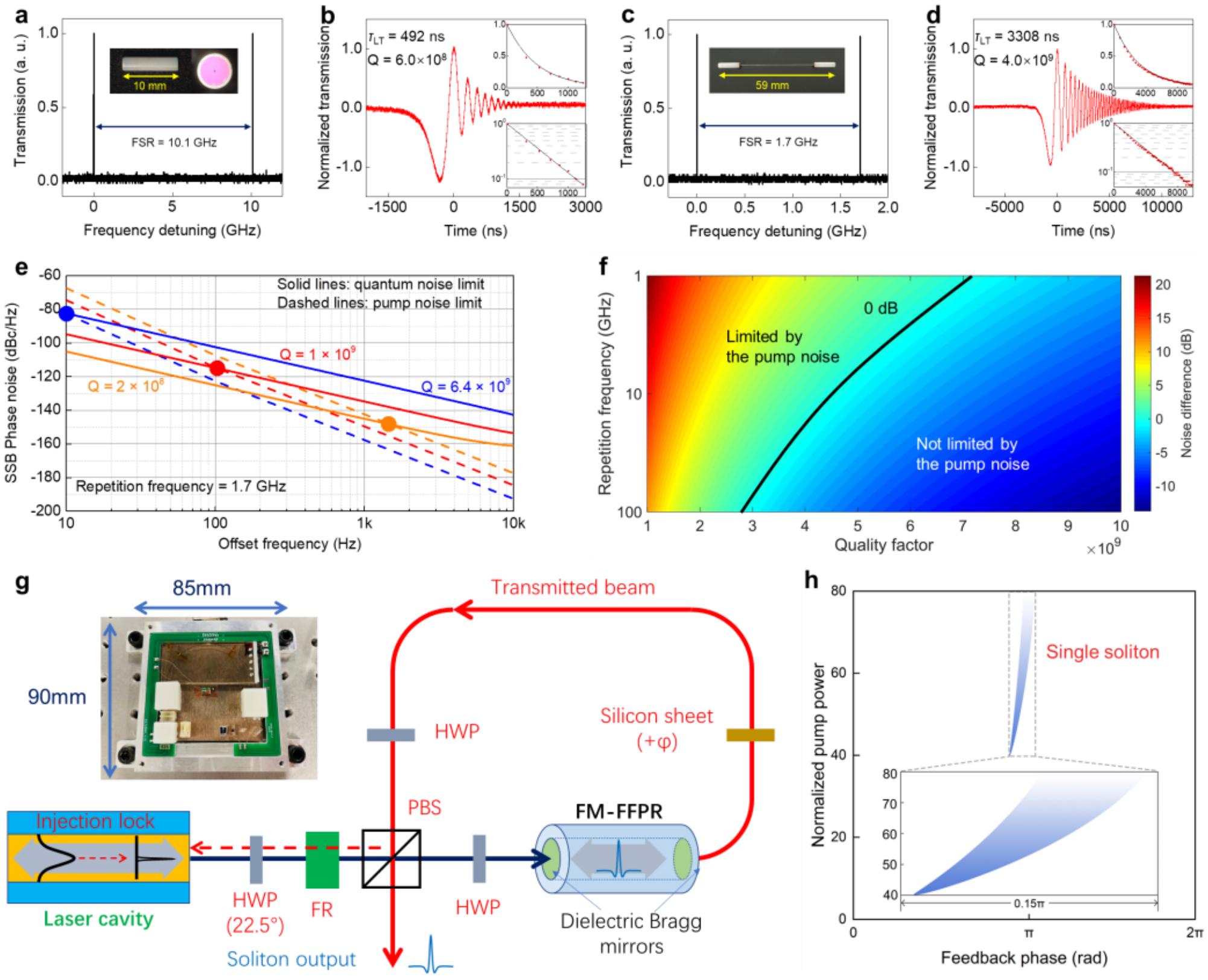

**Fig. 1 | Automated turnkey microcomb setup. a, c,** Cold-cavity transmission spectra around 1550 nm of 10.1 GHz and 1.7 GHz FFPRs, respectively. Insets: pictures of FFPRs and microscope photo of the end face. **b, d,** Ring-down traces of 10.1 GHz and 1.7 GHz FFPRs, respectively. Insets: fitted traces in linear and logarithmic scales, respectively. **e,** Theoretical quantum noise and pump noise limit of 1.7 GHz microcomb in FFPR with different Qs. The parameter used in the calculation is as follows: $Q_{LD} = 4 \times 10^3$, $\beta = 0.1$, and $\alpha_g = 2.5$[37,38,42]. $n_2 = 3\times10^{-20}$ m$^2$/W, effective mode field area $A_{eff} = 112$ μm$^2$, group velocity dispersion (GVD) $\beta_2 = -27$ fs$^2$/mm, $P_{in} = 100$ mW, $\lambda = 1545$ nm, $n = 1.45$. SSB, single sideband. **f,** Difference between the pump noise limit and quantum noise at 10 Hz offset frequency as a function of FFPR Q and repetition frequency. **g,** Self-injection locking soliton microcomb setup based on FFPR. The DFB laser's output is in horizontal polarization state. A set of half wave plate (HWP), Faraday rotator (FR) and polarizing beam splitter (PBS) effectively works as an optical circulator that blocks the back reflected beam from FFPR to be coupled into DFB laser cavity. A proportion of the transmitted beam is tuned to be vertically polarized by an HWP and then coupled back into DFB laser cavity through the PBS, while the residual horizontally polarized beam transmits through the PBS as output. A silicon sheet is inserted in the feedback loop for precise thermo-optic phase control. Inset: photograph of the compact soliton microcomb generator package with a size of 85 mm * 90 mm * 25 mm. **h,** Simulated single soliton existence parametric space as a function of the feedback phase ($\phi$) and pump power, which is normalized to the parametric oscillation threshold.

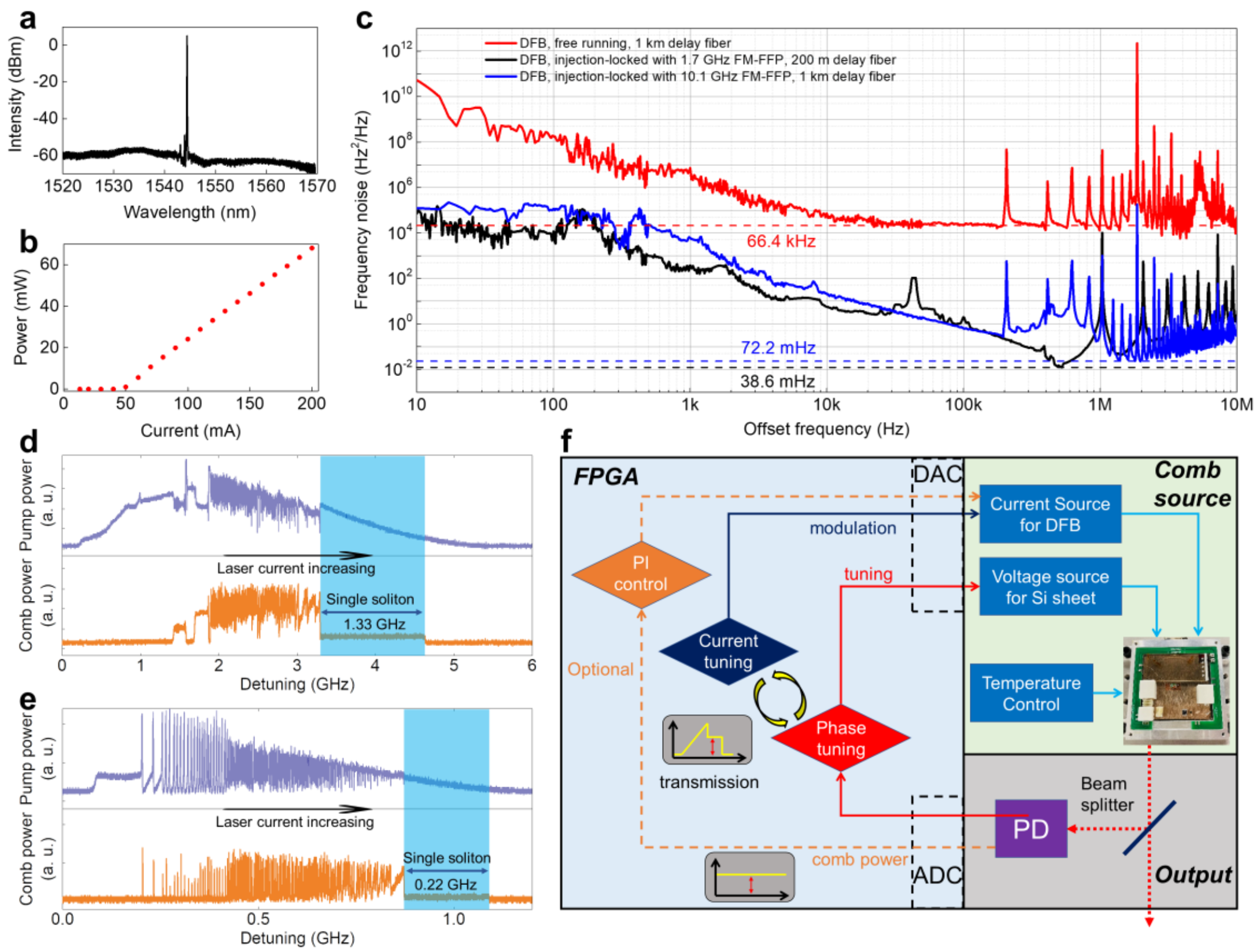

**Fig. 2 | Pump laser characterization and automated microcomb generation strategy.** **a,** Spectrum of the free-running DFB laser. **b,** DFB output power as a function of current. **c,** Frequency noise spectral densities of the DFB laser with free-running (red curve) and injection-locking to the FFPRs (black and blue curves). The dashed lines label the fundamental linewidth levels, indicating a noise reduction factor over $10^5$. **d, e,** 10.1 GHz and 1.7 GHz FFPRs transmission traces when sweeping DFB laser current with proper phase delay, respectively. Single soliton regions are as wide as 1.33 GHz and 0.22 GHz. **f,** The automated microcomb generation strategy based on FPGA.

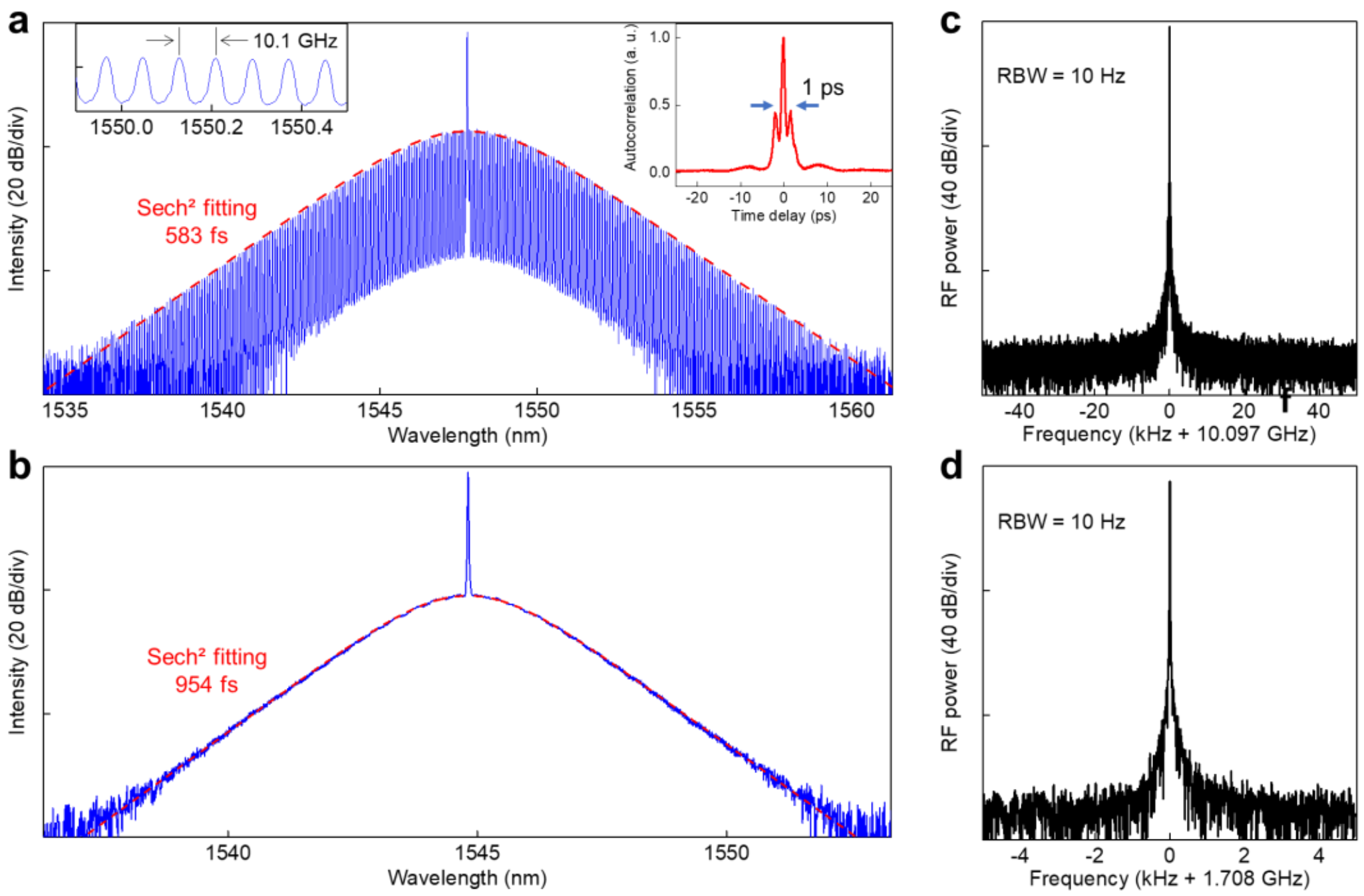

**Fig. 3 | Single soliton generation. a, b,** Optical spectra of single soliton states at 10.1 GHz and 1.7 GHz, respectively. Left inset of **a**: zoom-in spectrum of 10.1 GHz comb lines (OSA resolution is 0.02 nm). Right inset of **a**: autocorrelation trace of 10.1 GHz single soliton pulse. **c, d,** RF beatnotes of single soliton states at 10.1 GHz and 1.7 GHz, respectively. RBW, resolution bandwidth.

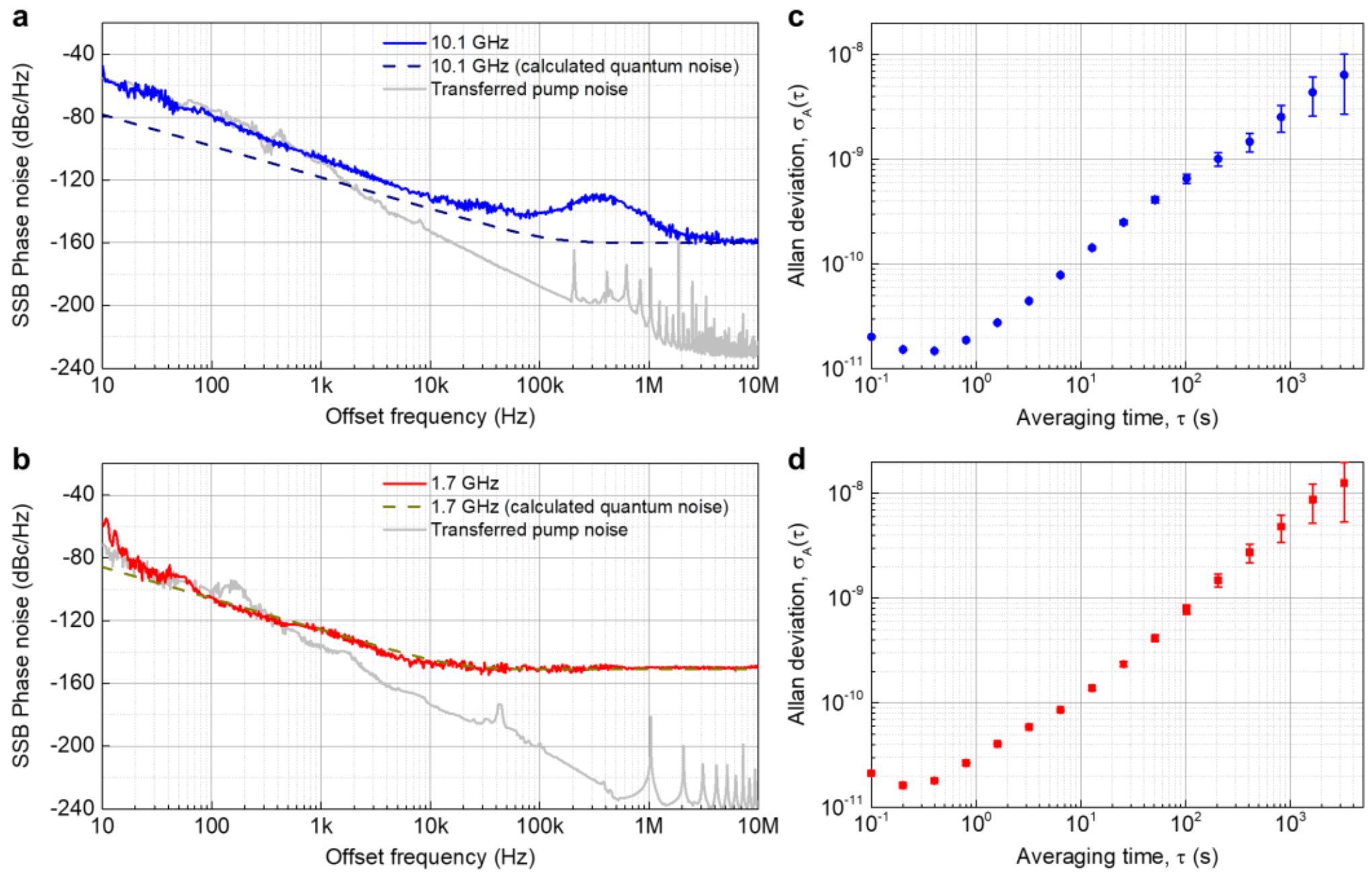

**Fig. 4 | Noise and stability characterization. a, b,** Single sideband (SSB) phase noises of the soliton repetition frequencies. The dashed lines represent for the calculated quantum noises. The gray lines represent for the noise transferred from injection-locked pump lasers, whose frequency noises are shown in Fig. 2c. The noise peaks that appear at 100~200 Hz offset frequency of the pump noise curve are likely to be caused by the vibration of the delay fiber in the heterodyne measurement system. **c, d,** Allan deviation of the 10.1 GHz and 1.7 GHz soliton repetition frequencies, respectively, measured with 0.1 s gate time.

**Acknowledgements:** We thank Prof. Fei Xu for providing the autocorrelator. We thank Dr. Bowen Li for fruitful discussion. This work was supported by National Key R&D Program of China (2019YFA0705000, 2017YFA0303700), Key R&D Program of Guangdong Province (2018B030329001), Leading-edge technology Program of Jiangsu Natural Science Foundation (BK20192001), National Natural Science Foundation of China (51890861, 11690031, 11621091, 11674169, 62075233), Guangdong Major Project of Basic and Applied Basic Research, Zhangjiang Laboratory (ZJSP21A001), China Postdoctoral Science Foundation (2022M710068), Jiangsu Planned Projects for Postdoctoral Research Funds (2021K259B).

**Conflict of interests:** The authors declare no competing interests.

**Author contributions:** K. J. and Z. X. conceived the original idea and designed the experiment. K. J., X. W., W. L. and Y. L. prepared the FFPRs sample, designed and packaged the microcomb engine including the DFB laser and FFPRs. K. J. and X. Y. performed the measurement and simulation, and conducted the data analysis. K. J., W. L. and Z. X. wrote the manuscript. Z. X. and S. Z. supervised the whole work. All authors contributed to the manuscript preparation.

**Supplementary information** is available for this paper.

**Correspondence and requests for materials** should be addressed to Kunpeng Jia, Wei Liang and Zhenda Xie.

# Supplementary Information for
# Automated turnkey microcomb for low-noise microwave synthesis

Kunpeng Jia[1]†*, Xinwei Yi[1]†, Xiaohan Wang[1], Yunfeng Liu[2], Shu-Wei Huang[3], Xiaoshun Jiang[1], Wei Liang[2]*, Zhenda Xie[1]*, Shi-ning Zhu[1]

*[1] National Laboratory of Solid State Microstructures, School of Electronic Science and Engineering, College of Engineering and Applied Sciences, School of Physics, and Collaborative Innovation Center of Advanced Microstructures, Nanjing University, Nanjing 210093, China.*

*[2] Suzhou Institute of Nano-tech and Nano-bionics, Chinese Academy of Sciences, Suzhou 215123, China.*

*[3] Department of Electrical, Computer and Energy Engineering, University of Colorado Boulder, Boulder, Colorado 80309, USA.*

* E-mail: jiakunpeng@nju.edu.cn; wliang2019@sinano.ac.cn; xiezhenda@nju.edu.cn

† These authors contributed equally to this work.

**This Supplementary Information consists of the following sections:**

## S1. Quantum-limited phase noise of direct-diode pumped microcomb

We consider the quantum-limited phase noise of a microcomb with direct diode pump, which has a limited power budget. The quantum-limited phase noise can be written as[1]

$$L_{\Phi} = \frac{2\sqrt{2}\pi^3\hbar c^3 n_2}{Vn^2\lambda^2\gamma^2}\sqrt{\frac{P_{th}}{P_{in}(-D)}}\left[1+\frac{P_{th}(-D)}{P_{in}}\frac{\gamma^2}{96f^2}+\frac{1}{24}\frac{P_{in}(-D)}{P_{th}}\left(1+\frac{\pi^2 f^2}{\gamma^2}\right)^{-1}\frac{\gamma^2}{\pi^2 f^2}\right], \quad \text{(S1)}$$

where $P_{\mathrm{th}}$ and $P_{\mathrm{in}}$ are the parametric threshold power and the pump power, respectively. Mode-locked soliton state is normally achieved at certain ratio $\kappa$ for $P_{\mathrm{in}}$ relative to $P_{\mathrm{th}}$. $D$, $\gamma$, $n_2$, $n$, $V$, $c$, $\lambda$, and $f$ are the dimensionless dispersion parameter, half width half maximum of the cavity resonance, nonlinear refractive index, refractive index, mode volume of the microresonator, speed of light in the vacuum, pump wavelength, and the offset radio frequency, respectively. Under critical coupling assumption, the relationship between parametric threshold power $P_{\mathrm{th}}$ and $\gamma$, $n_2$, $V$ is given by[2-4]

$$P_{th} = \frac{n^2 V\gamma^2\lambda}{2c^2 n_2}, \quad \text{(S2)}$$

so that Eq. (S1) can be rewritten as

$$L_{\Phi} = \frac{\sqrt{2}\pi^3\hbar c}{\lambda}\sqrt{\frac{P_{th}}{P_{in}(-D)}}\frac{1}{P_{th}}\left[1+\frac{P_{th}(-D)}{P_{in}}\frac{\gamma^2}{96f^2}+\frac{1}{24}\frac{P_{in}(-D)}{P_{th}}\left(1+\frac{\pi^2 f^2}{\gamma^2}\right)^{-1}\frac{\gamma^2}{\pi^2 f^2}\right]. \quad \text{(S3)}$$

Then we can get the Eq. (1) in the maintext where different terms in Eq. (S3) dominate in the low and high offset frequency ranges

$$\begin{aligned} L_{\Phi} &= \frac{\sqrt{2}\pi^3\hbar c}{\lambda}\sqrt{\frac{1}{\kappa(-D)}}\bullet\frac{1}{P_{th}}, \quad f \gg \gamma \\ L_{\Phi} &= \frac{\sqrt{2}\pi^5\hbar c^3}{\lambda^3}\sqrt{\frac{-D}{\kappa}}\left[\frac{1}{96\kappa}+\frac{\kappa}{24\pi^2}\right]\frac{1}{f^2}\bullet\frac{1}{P_{th}Q^2}, \quad f \ll \gamma \end{aligned} \quad \text{(S4)}$$

## S2. Numerical simulations of self-injection locking soliton microcomb generation

In this section we simulate the self-injection locking soliton microcomb generation in our fiber Fabry-Perot resonator (FFPR) system. The FFPR we developed has a different geometric configuration from traveling-wave resonators such like microring resonator and whispering gallery mode resonator. The self-injection locking based on traveling-wave resonators is accomplished by injecting the backscattering field of the resonators back into the laser cavity along the incident path[5-7]. Here in our scheme with FFPR, the transmission signal is utilized for feedback instead of the back reflection as illustrated in Fig. 1G of the main text. The power and phase delay of the feedback field can be adjusted independently. Meanwhile, the forward and backward propagating field in FFPR will introduce a nonlinear integral term into the Lugiato-Lefever equation (LLE)[8]. The complete equations of motions are:

$$\begin{aligned}\frac{\partial \psi}{\partial t} &= -(\kappa + i\delta\omega)\psi + i\kappa\psi|\psi|^2 + i\frac{L^2 D_2}{2\pi^2}\frac{\partial^2 \psi}{\partial z^2} + \frac{i\kappa}{L}\psi\int_{-L}^{L}|\psi|^2 d\theta - \sqrt{T_1\kappa\kappa_L}e^{i\phi_B}\psi_L \\ \frac{d\psi_L}{dt} &= i(\delta\omega_L - \delta\omega)\psi_L - \frac{\gamma}{2}\psi_L + \frac{g_L}{2}(1+i\alpha_g)\psi_L - \sqrt{T_2\kappa\kappa_L}e^{i\phi_B}\frac{1}{2\pi}\int_{-L}^{L}\psi d\theta\end{aligned} \quad \text{(S5)}$$

Here $\psi$ is the normalized soliton field amplitude, $\psi_L$ is normalized laser field amplitude, t is the evolution time variables, z is the resonator spatial variables. $\delta\omega$ is the detuning of the cold-cavity resonance compared to injection-locked laser. $\delta\omega_L$ is the detuning of the cold-cavity resonance relative to the cold laser frequency. $2\kappa = \frac{\omega}{Q}$ is the fullwidth-at-half-maximum resonator linewidth. $D_2 = \left.\frac{\partial^2 \omega_u}{\partial u^2}\right|_{u=0}$ is the second-order dispersion parameter. $2L$ is the roundtrip length of the resonator. $\gamma$ is the loss rate of laser mode. $g_L$ is the intensity dependent gain of the laser. $\alpha_g$ is the linewidth enhancement Henry factor[9,10]. $\phi_B$ is the adjustable phase delay between resonator and laser cavity. $T_1$ is the laser cavity-to-resonator coupling efficiency. $T_2$ is the resonator-to-laser cavity feedback efficiency.

The proposed theoretical model of Fabry-Perot self-injection locking can be further normalized to facilitate discussion. By normalizing the time variables t and the spatial variables z to be $\tau = \kappa t$ and $\theta = z\frac{\pi}{L}$, respectively, we have

$$\begin{aligned}\frac{\partial \psi}{\partial \tau} &= -\psi - i\frac{\delta\omega}{\kappa}\psi + i\psi|\psi|^2 + i\frac{D_2}{2\kappa}\frac{\partial^2 \psi}{\partial \theta^2} + i\frac{1}{\pi}\psi\int_{-\pi}^{\pi}|\psi|^2 d\theta - \frac{\sqrt{T_1\kappa\kappa_L}}{\kappa}e^{i\phi_B}\psi_L \\ \frac{d\psi_L}{d\tau} &= i\frac{\delta\omega_L - \delta\omega}{\kappa}\psi_L - \frac{\gamma}{2\kappa}\psi_L + \frac{g_L(1+i\alpha_g)}{2\kappa}\psi_L - \frac{\sqrt{T_2\kappa\kappa_L}}{\kappa}e^{i\phi_B}\frac{1}{2\pi}\int_{-\pi}^{\pi}\psi d\theta\end{aligned} \quad \text{(S6)}$$

By defining normalized detuning of resonator as $\alpha = \frac{\delta\omega}{\kappa}$, normalized detuning of laser as $\alpha_L = \frac{\delta\omega_L + \frac{\gamma}{2}\alpha_g}{\kappa}$, normalized second-order dispersion parameter as $\beta = -\frac{D_2}{\kappa}$, normalized pump input power as $F = \sqrt{\frac{\kappa_L}{\kappa} T_1} |\psi_L|$, the equations can be arranged as

$$\frac{\partial \psi}{\partial \tau} = -\psi - i\alpha\psi + i\psi|\psi|^2 - i\frac{\beta}{2}\frac{\partial^2 \psi}{\partial \theta^2} + i\frac{1}{\pi}\psi\int_{-\pi}^{\pi}|\psi|^2 d\theta - e^{i\phi_B}e^{i\phi_L}F$$
$$\frac{d\psi_L}{d\tau} = i(\alpha_L - \alpha)\psi_L - \frac{\gamma}{2\kappa}\psi_L + \frac{g_L(1+i\alpha_g)}{2\kappa}\psi_L - \sqrt{\frac{T_2}{T_1}}\frac{1}{|\psi_L|}e^{i\phi_B}F\frac{1}{2\pi}\int_{-\pi}^{\pi}\psi d\theta \tag{S7}$$

The laser field can be decomposed into amplitude and phase part, which are

$$\frac{1}{|\psi_L|}\frac{d|\psi_L|}{d\tau} = -\frac{\gamma}{2\kappa} + \frac{g_L}{2\kappa} - \mathrm{Re}[\sqrt{\frac{T_2\kappa_L}{\kappa}}e^{i\phi_B}\frac{\int_{-\pi}^{\pi}\psi d\theta}{2\pi\psi_L}]$$
$$\frac{d\phi_L}{d\tau} = (\alpha_L - \alpha) + \frac{g_L}{2\kappa}\alpha_g - \mathrm{Im}[\sqrt{\frac{T_2\kappa_L}{\kappa}}e^{i\phi_B}\frac{\int_{-\pi}^{\pi}\psi d\theta}{2\pi\psi_L}] \tag{S8}$$

As the laser amplitude variation is much faster than Kerr comb generation dynamics in resonator, we can assume its variation as an instantaneous process (i.e., assume $\frac{d|\psi_L|}{d\tau} = 0$). The intensity dependent gain of the laser $g_L$ can be solved as

$$\frac{g_L}{2} = \frac{\gamma}{2} + \mathrm{Re}[\sqrt{T_2\kappa\kappa_L}e^{i\phi_B}\frac{\int_{-\pi}^{\pi}\psi d\theta}{2\pi\psi_L}], \tag{S9}$$

where Re[•] and Im[•] are the real and imaginary part functions, respectively. Substituting $\Phi = -e^{i\phi_L}e^{i\phi_B}$ and Eq. (S8) into the phase part of the normalized laser field equation in Eq. (S7), we get

$$\frac{\partial \psi}{\partial \tau} = -(1+i\alpha)\psi + i\psi|\psi|^2 - i\frac{\beta}{2}\frac{\partial^2 \psi}{\partial \theta^2} + i\frac{1}{\pi}\psi\int_{-\pi}^{\pi}|\psi|^2 d\theta + \Phi F$$
$$\frac{1}{i\Phi}\frac{d\Phi}{d\tau} = \alpha_L - \alpha - \mathrm{Im}[\sqrt{1+a_g^2}e^{i(\phi_B - \phi_L - \arctan(\alpha_g) + \frac{\pi}{2})}\frac{\kappa_L\sqrt{T_1T_2}}{\kappa}\frac{\int_{-\pi}^{\pi}\psi d\theta}{2i\pi F}] \tag{S10}$$

Finally, by defining locking strength $K = \frac{\kappa_L}{\kappa}\sqrt{T_1T_2}\sqrt{1+\alpha_g^2}$ and derived feedback phase $\phi = 2\phi_B - \arctan(\alpha_g) + \frac{\pi}{2}$, we obtain normalized simulation equations

$$\frac{\partial \psi}{\partial \tau} = -(1+i\alpha)\psi - i\frac{\beta}{2}\frac{\partial^2 \psi}{\partial \theta^2} + 2i\psi\frac{1}{2\pi}\int_{-\pi}^{\pi}|\psi|^2 d\theta + i\psi|\psi|^2 + \Phi F$$
$$\frac{1}{i\Phi}\frac{d\Phi}{d\tau} = \alpha_L - \alpha + K\,\mathrm{Im}[e^{i\phi}\frac{\int_{-\pi}^{\pi}\psi d\theta}{2i\pi\Phi F}] \tag{S11}$$

For stationary solutions when the field in resonator reaches a steady-state $\psi_S$, the equations can be reduced to be

$$
\begin{aligned}
&-(1+i\alpha)\psi_S + 2i\psi_S\left|\psi_S\right|^2 + i\psi_S\left|\psi_S\right|^2 + \Phi F = 0 \\
&\alpha = \alpha_L + K\,\mathrm{Im}[e^{i\phi}\frac{\psi_S}{i\Phi F}]
\end{aligned}
\tag{S12}
$$

with $\rho = \frac{1}{2\pi}\int_{-\pi}^{\pi}\psi_S d\theta$ and $P = \frac{1}{2\pi}\int_{-\pi}^{\pi}\left|\psi_S\right|^2 d\theta$. The locking equilibrium of normalized detuning $\alpha$ can be obtained by combining these two equations

$$
\alpha = \alpha_L - K(\frac{\cos\phi + \sin\phi(\alpha - P)}{1+(\alpha-3P)^2}). \tag{S13}
$$

To obtain analytic solution of $\alpha$, we set $\frac{\cos\phi + \sin\phi(\alpha - P)}{1+(\alpha-3P)^2} = 0$ then we get[5]

$$
\alpha = 3P - \cot(\phi). \tag{S14}
$$

We mainly interest in the generation of single soliton, which can be obtained with appropriate pump power ($F$), feedback phase ($\phi$) and laser detuning ($\alpha_L$) as shown in Fig. S1. While simulation begins, the field in resonator grows rapidly from background noise to modulation instability (MI) combs and finally reaches single soliton state. Figure 1H of the main text gives the single soliton existence space of pump power and feedback phase in our microcomb system.

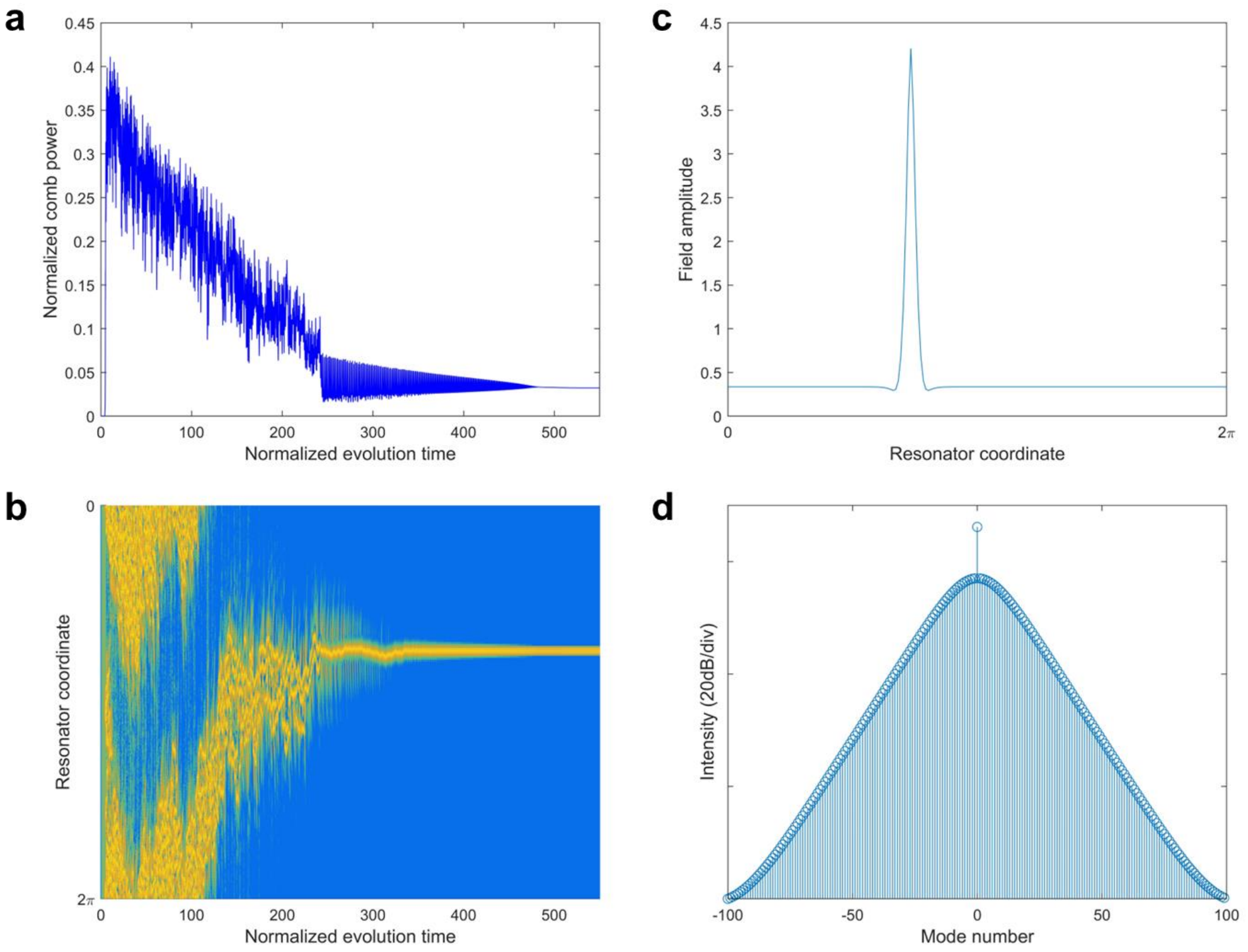

**Fig. S1 | Numerical simulations of self-injection locking soliton microcomb**

**generation with FFPR.** Parameters are $\beta = -0.0229$, $K = 16819$, $F^2 = 81$, $\phi = \frac{257}{256}\pi$ and $\alpha_L = -840$. **a, b,** Soliton field power and coordinate amplitude evolution with time, respectively. **c, d,** The soliton field amplitude in time domain and soliton spectrum at evolution time $\tau = 550$.

## S3. Other comb states.

### S3.1. Multiple soliton state.

By adjusting the soliton recognizer algorithm, different multiple soliton states can also be stably accessed. The spectra of several multiple soliton states are shown in Figs. S3 and S3.

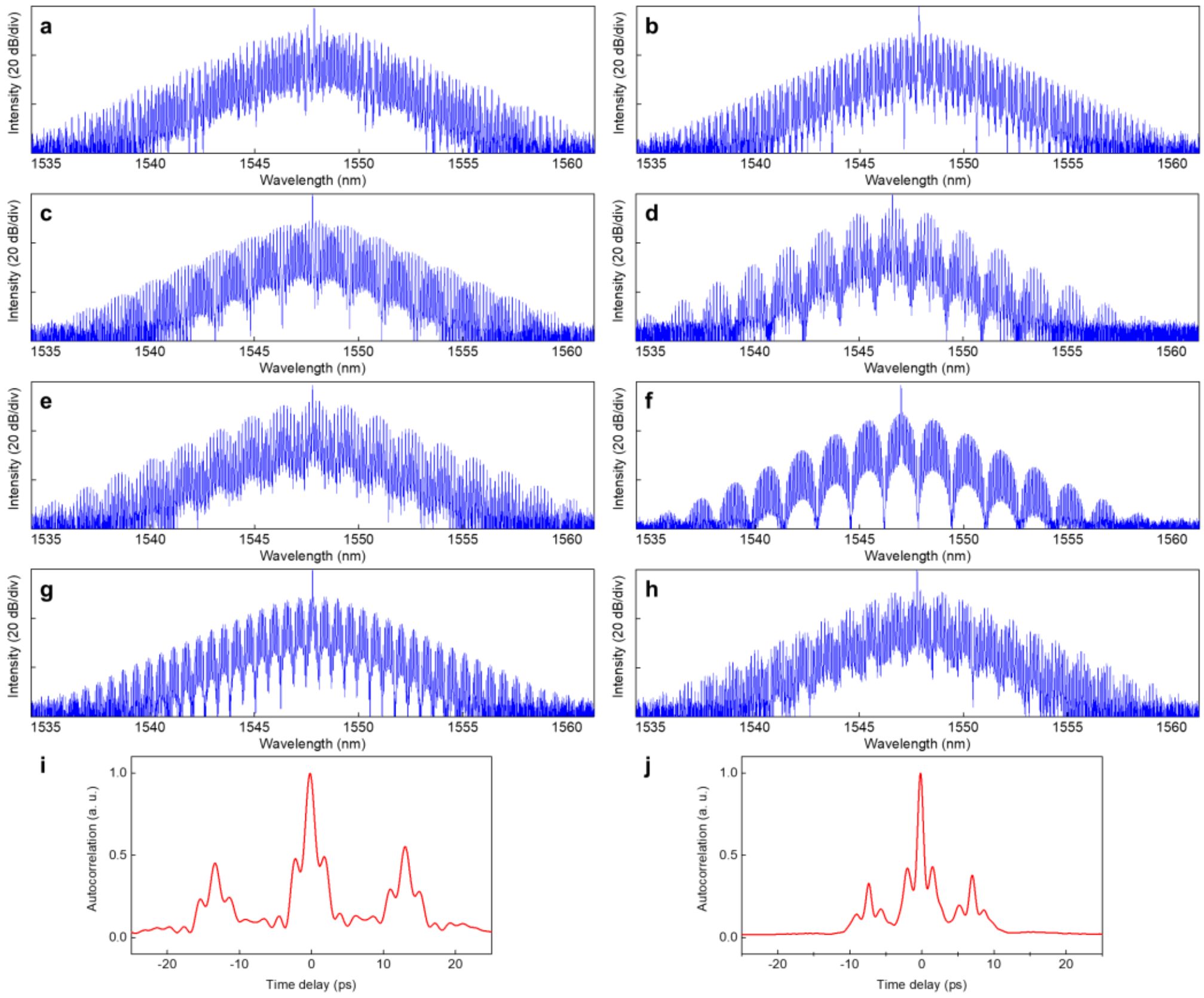

**Fig. S2 | Multiple soliton states generated in 10.1 GHz FFPR. a-h,** spectra of multiple soliton states. **i, j,** autocorrelation traces of multiple soliton states in **g** and **h**, respectively.

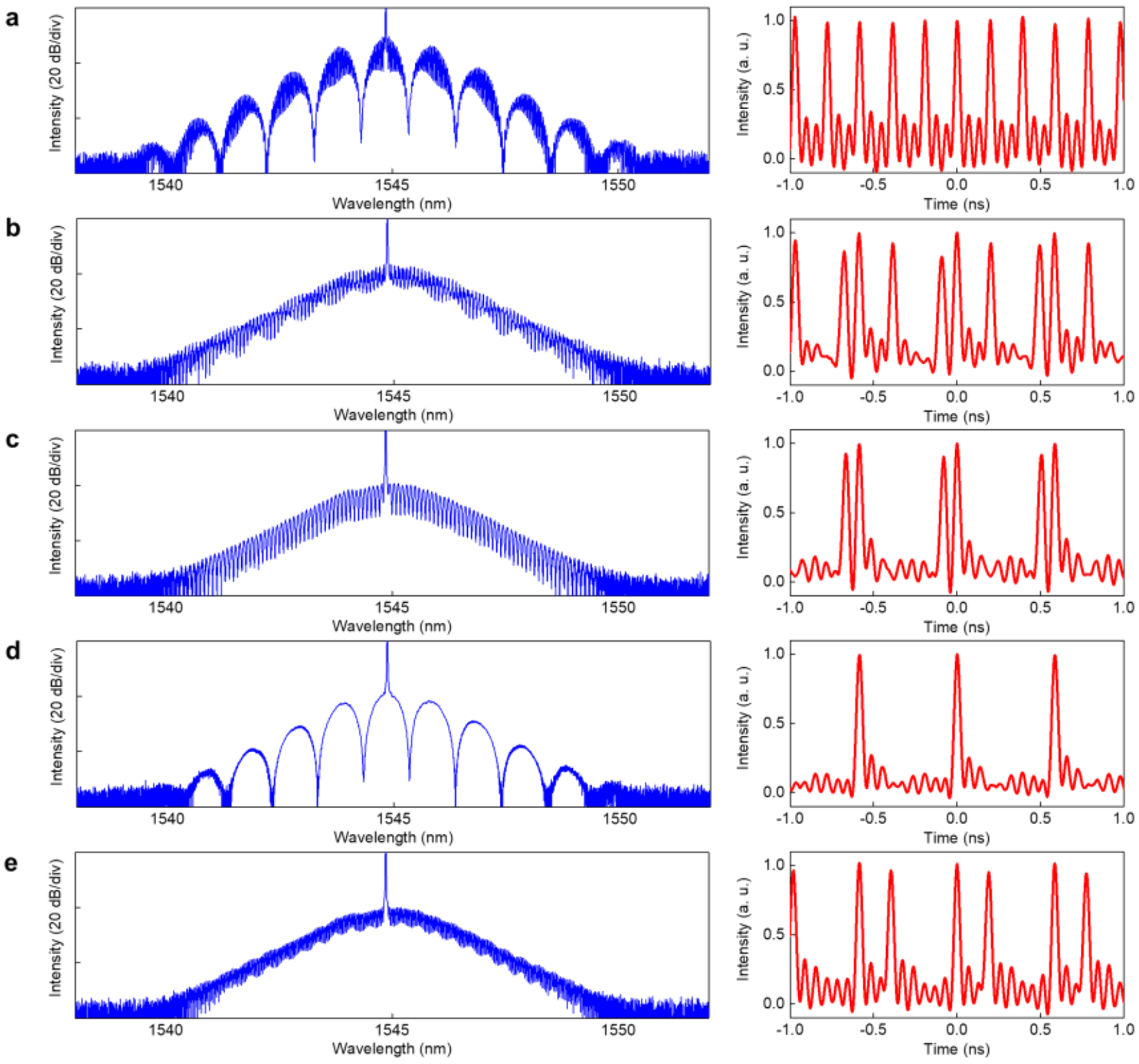

**Fig. S3 | Multiple soliton states generated in 1.7 GHz FFPR.** Left panels: spectra of multiple soliton states. Right panels: pulse trains measured by a fast-speed oscilloscope.

## S3.2. Soliton crystal.

In addition to the capability for microcombs generation at a few gigahertz, FFPRs also support high-coherence microcomb states with larger comb spacing over tens of gigahertz such like soliton crystal state. Figure S4 plots two different soliton crystal states with multi-FSR (free spectral range) comb spacings. The corresponding photodetections of the low-intensity noise from the comb lines beating indicate high coherence.

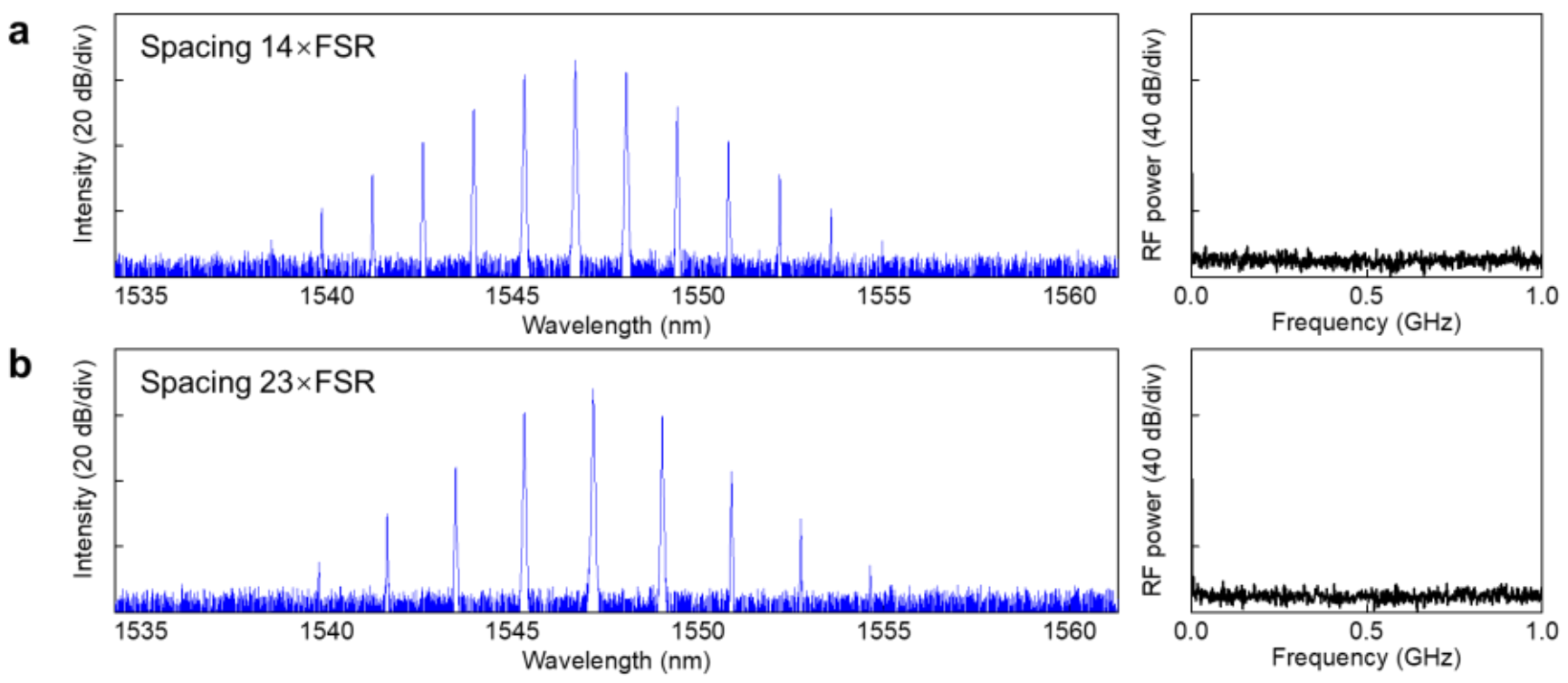

**Fig. S4 | Soliton crystal states generated in 10.1 GHz FFPR.** Left panels: spectra of soliton crystal states with comb spacings of 140 GHz and 230 GHz, respectively. Right

panels: RF signals showing low-intensity noise.

### S3.3. Breather soliton.

By slightly increasing the laser frequency when a single or multiple soliton state is generated, we can stably access breather soliton states. Breather soliton is a special soliton state whose energy localized in space with temporal oscillations (or vice versa), as shown in the left panels of Fig. S5a. The intensity envelope of breather soliton oscillates with time. The right panels of Fig. S5b plot the corresponding RF signals at repetition frequency.

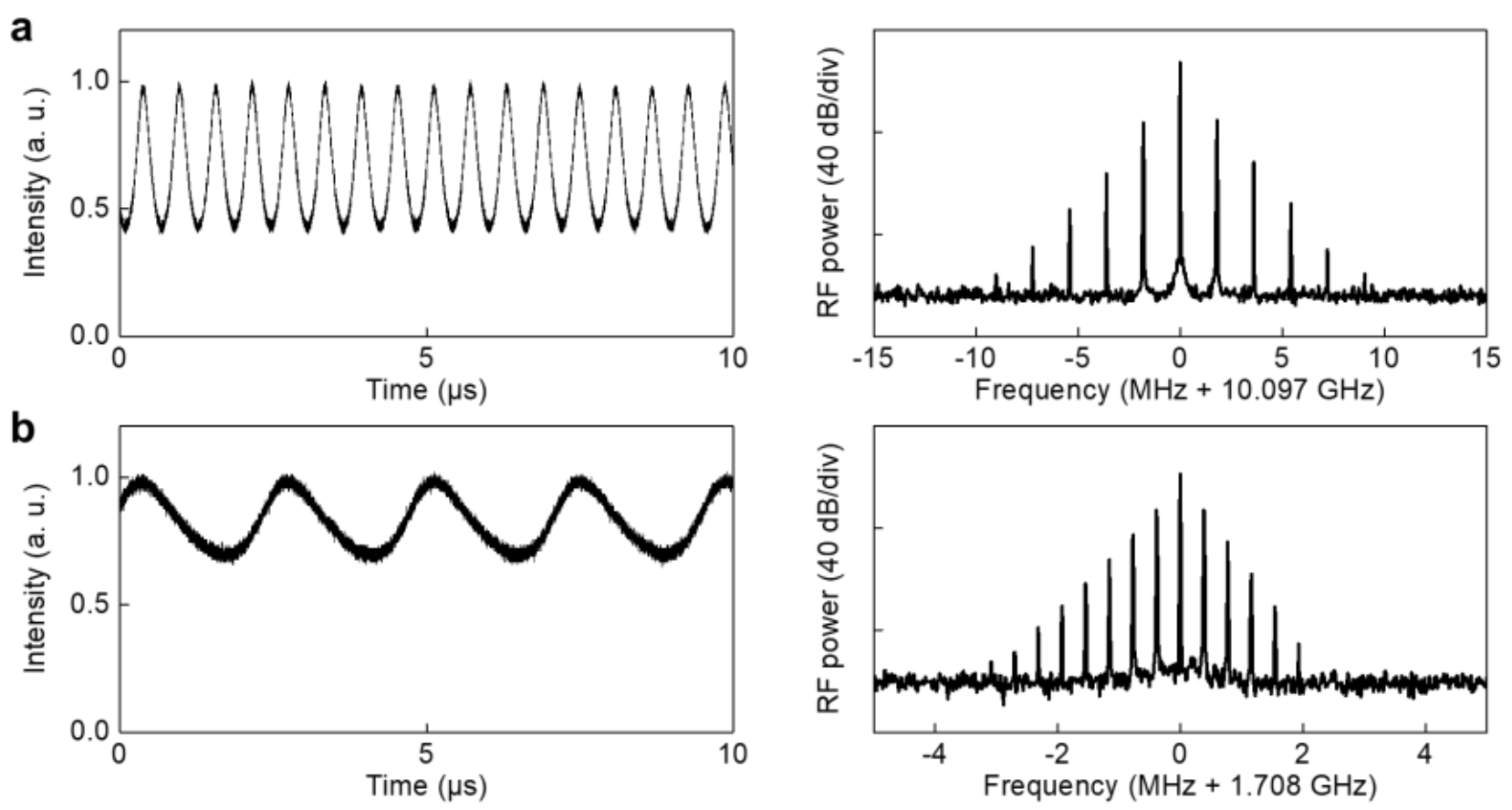

**Fig. S5 | Breather soliton states.** Left panels: time domain signals. Right panels: RF signals. **a,** 10.1 GHz microcomb. **b,** 1.7 GHz microcomb.

## Supplementary references